%% file: main.tex
\documentclass[sigconf]{acmart-me}

\usepackage{booktabs} 
\usepackage{url}
\usepackage{color}
\usepackage{enumitem}
\setcopyright{rightsretained}

\acmDOI{}

\acmISBN{}

\acmConference[MediaEval'19]{Multimedia Evaluation Workshop}{27-29 October 2019}{Sophia Antipolis, France} 
\acmYear{}
\copyrightyear{}

\acmPrice{}

\begin{document}
\title{Emotion and Theme Recognition in Music \\
with Frequency-Aware RF-Regularized CNNs}

\author{Khaled Koutini, Shreyan Chowdhury, Verena Haunschmid, Hamid Eghbal-zadeh, Gerhard Widmer}
\affiliation{Johannes Kepler University Linz\\ }
\email{firstname.lastname@jku.at}

%
%
%
%
%

\renewcommand{\shortauthors}{K. Koutini et al.}
\renewcommand{\shorttitle}{Emotion and Theme Recognition in Music}

\begin{abstract}
We present CP-JKU submission to MediaEval 2019; a Receptive Field-(RF)-regularized and Frequency-Aware CNN approach for tagging music with emotion/mood labels. We perform an investigation regarding the impact of the RF of the CNNs on their performance on this dataset. We observe that  ResNets with smaller receptive fields -- originally adapted for acoustic scene classification -- also perform well in the emotion tagging task. We improve the performance of such architectures using techniques such as Frequency Awareness and Shake-Shake regularization, which were used in previous work on general acoustic recognition tasks.

\end{abstract}

%
%
%
%
%


\maketitle
\section{Introduction}
\label{sec:intro}

\input{intro.tex}

\section{Setup}
\label{sec:setup}

\input{setup.tex}





\section{Adapting CNNs}
\input{appro.tex}

\label{sec:approach}

\section{Submissions and Results}

\input{res.tex}

\begin{acks}
This work has been supported by the LCM – K2 Center within the framework of the Austrian COMET-K2 program,
and the European Research Council (ERC) under the EU's Horizon 2020 research and innovation programme, under grant agreement No 670035 (project ``Con Espressione'').
\end{acks}

\bibliographystyle{ACM-Reference-Format}
\def\bibfont{\small} 
\bibliography{sigproc} 

\end{document}

%% file: intro.tex
Content based emotion recognition in music is a challenging task in part because of noisy datasets and unavailability of royalty-free audio of consistent quality. 
The recently released MTG-Jamendo dataset \cite{bogdanov2019mtg} is aimed at addressing these issues. 

The Emotion and Theme Recognition Task of MediaEval 2019 uses a subset of this dataset containing relevant emotion tags, and the task objective is to predict scores and decisions for these tags from audio (or spectrograms). 
The details of this specific data subset, task description, data splits, and evaluation strategy can be found in the overview paper \cite{overview}.

Convolutional Neural Networks (CNNs) achieve state-of-the-art results in many tasks such as image classification~\cite{heDeepResidualLearning2016,huangDenselyConnectedConvolutional2017}, acoustic scene classification~\cite{DorferDCASE2018task1,Koutinitrrfcnns2019} and audio tagging~\cite{DorferDCASE2018task2}.
These models can learn their own features and classifiers in an end-to-end fashion, which as a result reduces the need for task-specific feature engineering.
Although CNNs are capable of learning high-level concepts given very simple and low-level information, the careful design of the network architectures in CNNs is a crucial step in achieving good results.

In a recent study~\cite{Koutini2019Receptive,Koutinitrrfcnns2019}, Koutini et al. showed that the \emph{receptive field (RF)} of CNN architectures is a very important factor when it comes to processing audio signals.
Based on these findings, a regularization technique was proposed, that can significantly boost the performance of CNNs when used with spectrogram features.
Further, in~\cite{koutinifaresnet2019} a drawback of CNNs in the audio domain is highlighted, which is caused by the lack of spatial ordering in convolutional layers. As a solution, \emph{Frequency-Aware (FA) Convolutional Layers} were introduced, to be used in CNNs with the commonly-used spectrogram input.


The proposed RF-regularization and FA-CNNs have shown great promise in several tasks in the field of Computational Auditory Scene Analysis (CASA), and achieved top ranks in international challenges~\cite{Koutinitrrfcnns2019}.
In this report, we extend the previous work 
to Music Information Retrieval (MIR) and demonstrate that these models can be used to recognize emotion in music, and achieve new state-of-the-art results.

%% file: setup.tex
\subsection{Data Preparation}
\label{sec:setup:dataprep}
We used a sampling rate of 44.1 kHz to extract the input features. We apply a Short Time Fourier
Transform (STFT). The window size for the STFT is 2048 samples and the overlap between windows is  75\% for submissions 1, 2 and 3, and 25\% for submissions 4 and 5. We use perceptually-weighted Mel-scaled spectrograms similar to ~\cite{DorferDCASE2018task1,Koutini2019Receptive,Koutinitrrfcnns2019}, which results in an input having 256 Mel bins in the  frequency dimension. 

\subsection{Optimization}
In a setup similar to ~\cite{Koutini2019Receptive,Koutinitrrfcnns2019,koutinifaresnet2019}, we use Adam~\cite{kingmaAdamMethodStochastic2014} for 200 epochs. We start with 10 epochs warm-up learning rate, we train with a constant learning rate of  $1 \times 10^{-4}$ for 60 epochs. After that, we use a linear learning rate scheduler for 50 epochs, dropping the learning rate to $1 \times 10^{-6}$. We finally train for 80 more epochs using the final learning rate.

\subsection{Data Augmentation}
\textit{Mix-up}~\cite{zhangMixupEmpiricalRisk2017} has proven essential in our experiments to boost the perfomance and the generalization of our models. These results are consistent with experience from our previous work ~\cite{Koutini2019Receptive,Koutinitrrfcnns2019,koutinifaresnet2019}.

%% file: appro.tex

Convolutional Neural Networks (CNNs) have shown great success in many acoustic tasks~\cite{eghbal-zadehCPJKUSubmissionsDCASE20162016,hersheyCNNArchitecturesLargescale2017,lehnerClassifyingShortAcoustic2017,DorferDCASE2018task1,SakashitaDCASE2018task1,DorferDCASE2018task2,IqbalDCASE2018task2,LeeDCASE2017task4,KoutiniDCASE2018task4,Koutinitrrfcnns2019,koutinifaresnet2019,Koutini2019Receptive}. In our submissions, we build on this success and investigate their performance on tasks more specific to music. We use mainly adapted versions of ResNet~\cite{heDeepResidualLearning2016}. We adapt the architectures to the task using the guidelines proposed in Koutini et al.\cite{Koutini2019Receptive}\footnote{The source code is published at \url{https://github.com/kkoutini/cpjku_dcase19}}. We use the CNN variants introduced in~\cite{koutinifaresnet2019}.

\subsection{Receptive Field Regularization}
\label{sec:rf_regularization}
Limiting the receptive field (RF) has been shown to have a great impact on the performance of a CNN in a number of acoustic recognition and detection tasks~\cite{Koutini2019Receptive,Koutinitrrfcnns2019}. We investigated the influence of the receptive field in this task in a setup similar to ~\cite{Koutini2019Receptive}.


\begin{figure}[h]
  \centering
  \centerline{\includegraphics[width=0.9\columnwidth]{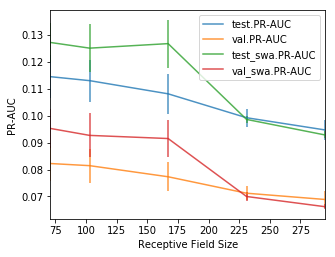}}
  \caption{PR-AUC for ResNets with different RFs}
  \label{fig:rf_exp}
\end{figure}

Figure~\ref{fig:rf_exp} shows the PR-AUC on both the the validation (val) and testing (test) sets, for ResNet models with different receptive fields and their SWA (see Section~\ref{sec:swa} below) variants. The results show the larger receptive field causes performance drops in accordance to the findings of ~\cite{Koutini2019Receptive}. Moreover, further experiments showed that size of the receptive field over the time dimension has lower significance on performance. 

\subsection{Frequency-Awareness and FA-ResNet}
Figure~\ref{fig:rf_exp} shows that smaller-RF ResNets perform better. As shown in~\cite{koutinifaresnet2019}, Frequency-Awareness can compensate for the lack of freuqency information caused by the smaller RF. We use Frequency-Aware ResNet (FA-ResNet) introduced in~\cite{koutinifaresnet2019}.

\subsection{Shake-Shake Regularization}

The Shake-Shake regularization~\cite{gastaldi2017shake} is proposed for improved stability and robustness. As shown in ~\cite{Koutinitrrfcnns2019} and~\cite{koutinifaresnet2019}, although Shake-Shake ResNets do not perform well in the original acoustic scene classification problem, it performed really well in this task.


\subsection{Model Averaging}
\label{sec:swa}
\textbf{Stochastic Weight Averaging:} Similar to ~\cite{Koutinitrrfcnns2019,koutinifaresnet2019},
we use Stochastic Weight Averaging (SWA)~\cite{izmailov2018averaging}. We add networks weights to the average every 3 epochs. The averaged networks turned out to out-perform each of the single networks.
\\\textbf{Snapshot Averaging}: When computing the final prediction we also average the predictions of 5 snapshots of the networks during training. Specifically, we average the model with the highest PR-AUC on the validation set with the last 4 SWA models' predictions during training.
\\\textbf{Multi-model Averaging}: We average different models that have different architectures, initialization and/or receptive fields over time.

%% file: res.tex
\subsection{Submitted Models}
%
%
Overall, we submitted five models to the challenge: the first three are variations of the approach described above; the other two were models tested during our experiments, and were submitted as additional baselines against which to compare our modified CNNs.
\newline\textbf{ShakeFAResNet} We average the prediction of 5 Shake-Shake regularized FA-ResNets with different initlizations. Their frequency RF is regularized as explained in Section~\ref{sec:rf_regularization}. They have however different RF over the time dimension.
\newline\textbf{FAResNet} similar to Shakefaresnet, but without Shake-Shake regularization.
\newline\textbf{Avg\_ensemble} We average the prediction of all the models included in both Shakefaresnet and Faresnet. In addition, we add a RF-regularized ResNet and DenseNet as introduced in~\cite{Koutini2019Receptive}.
\newline\textbf{ResNet34} In our preliminary experiments, Vanilla Resnet-34 outperformed Resnet-18 and Resnet-50 on the validation set, so we picked this architecture as an additional baseline.
\newline\textbf{CRNN}
The CRNN network was motivated by the notion that global structure of musical features could affect the perception of certain aspects of music (like mood), as mentioned by Choi et al \cite{choi2017convolutional}. We use an architecture similar to the one used by Choi et al, where the CNN part acts as the feature extractor and the RNN part acts as a temporal aggregator. This approach increased the performance from the baseline CNN and the Resnet-34.
\newline\textbf{CP\_ResNet} (not submitted to the challenge) We also show the results of a single model RF-regularized ResNet.



\subsection{Results}

\begin{table}[]
\caption{PR-AUC results}
\label{tab:results}
\begin{tabular}{lll}
 \toprule
Submission    & \begin{tabular}[c]{@{}l@{}}Validation\\ PR-AUC\end{tabular} & \begin{tabular}[c]{@{}l@{}}Testing\\ PR-AUC\end{tabular} \\ 
\midrule
ShakeFAResNet* & .1132& .1480\\
FAResNet*      & .1149& .1463\\
Avg\_ensemble* & \textbf{.1189}& \textbf{.1546}\\
ResNet34      & .0924& .1021\\
CRNN          & .0924 & .1172\\
CP\_ResNet &\textbf{.1097}& \textbf{.1325}\\
VGG-ish-baseline &-& .1077 \\
popular baseline &-& .0319\\
\bottomrule

*: indicates an ensemble. \\

\end{tabular}
\end{table}

Table ~\ref{tab:results} shows the results of our submitted systems and compares them with the baselines. We can see that our RF-regularized and Frequency-Aware CNNs outperform the baselines by a significant margin, resulting in ranking as the top 3 submissions in the challenge. 
The systems that are marked with a star *, are ensembles of multiple models and snapshots (Section\ref{sec:swa}). Table ~\ref{tab:results} also shows a single RF-regularized ResNet (CP\_ResNet) can perform very well compared to the baselines.